\documentclass[%
 aip,
 amsmath,amssymb,
 reprint,%
]{revtex4-1}

\usepackage{graphicx}
\usepackage{dcolumn}
\usepackage{bm}

\usepackage[utf8]{inputenc}
\usepackage[T1]{fontenc}
\usepackage{mathptmx}
\usepackage{siunitx}
\usepackage{threeparttable}
\usepackage[dvipsnames]{xcolor}
\usepackage{color,soul}
\usepackage[normalem]{ulem}
\begin{document}

\preprint{AIP/123-QED}

\title[Spontaneous Electron Emission Versus Dissociation in Internally Hot Silver Dimer Anions]{Spontaneous Electron Emission Versus Dissociation \\in Internally Hot Silver Dimer Anions}
\author{P. Jasik}
\email{patryk.jasik@pg.edu.pl}
\affiliation{Faculty of Applied Physics and Mathematics, Gdańsk University of Technology, 80-233 Gdańsk, Poland}
\affiliation{BioTechMed Center, Gdańsk University of Technology, 80-233 Gdańsk, Poland}
\author{J. Franz}
\affiliation{Faculty of Applied Physics and Mathematics, Gdańsk University of Technology, 80-233 Gdańsk, Poland}
\affiliation{Advanced Materials Center, Gdańsk University of Technology, 80-233 Gdańsk, Poland}
\author{D. K{\k e}dziera}%
\affiliation{Faculty of Chemistry, Nicolaus Copernicus University, 87-100 Toruń, Poland}
\author{T. Kilich}
\author{J. Kozicki}
\author{J.E. Sienkiewicz}
\affiliation{Faculty of Applied Physics and Mathematics, Gdańsk University of Technology, 80-233 Gdańsk, Poland}
\affiliation{Advanced Materials Center, Gdańsk University of Technology, 80-233 Gdańsk, Poland}

\quad \\
\quad \\

\begin{abstract}
Referring to a recent experiment, we theoretically study the process of a two-channel decay of the diatomic silver anion (Ag$_2^-$), namely the spontaneous electron ejection giving Ag$_2$ + e$^-$ and the dissociation leading to Ag$^-$ + Ag. The ground state potential energy curves of the silver molecules of diatomic neutral and negative ion were calculated using proper pseudo-potentials and atomic basis sets. We also estimated the non-adiabatic electronic coupling between the ground state of Ag$_2^-$ and the ground state of Ag$_2$ + e$^-$, which in turn allowed us to estimate the minimal and mean values of the electron autodetachment lifetimes. The relative energies of the rovibrational levels allow the description of the spontaneous electron emission process, while the description of the rotational dissociation is treated with the quantum dynamics method as well as time-independent methods. The results of our calculations are verified by comparison with experimental data.
\end{abstract}

\maketitle

\section{\label{sec:level1} Introduction}

The problem of coupling nuclear and electronic motions is still an area where much work needs to be done\cite{Sutcliffe2021}. Solving this problem fully nonadiabatically, without the separation of electronic and nuclear motions, is challenging even for the simplest molecule H$_2$, although for hydrogen isotopologues it has been solved with extraordinary accuracy\cite{Pachucki2016} including both relativistic\cite{Puchalski2018} and QED\cite{Puchalski2019} effects. Currently, less accurate nonadiabatic calculations can be done for slightly heavier molecules, such as few-electron diatomics\cite{Scheu2001,Bubin2009}. Treating electrons and nuclei on the same footing seems rather impossible for many-electron polyatomic molecules, but there is significant progress in incorporating nuclear quantum effects and non-Born-Oppenheimer effects, for specified nuclei, into calculations using nuclear-electronic orbital approach\cite{Pavosevic2020}. Nevertheless, most of the progress is made using the well-established concept of interatomic potentials, usually calculated using the Born-Oppenheimer approximation with added nonadiabatic couplings between different electronic states\cite{Reimers2015}. This approach has proven to be quite effective in describing ultrafast reorganization of electronic density during and after excitation by intense ultra-short electromagnetic pulses\cite{Kling2008}. This kind of process should be studied in the attosecond regime. Already, a great deal of effort is being put into the experimental and theoretical research which is leading to the emergence of a new field called attosecond science\cite{Corkum2007}.

Quite the opposite situation takes place, when the nonadiabatic coupling is minimal, which is the case when the Born-Oppenheimer potential curves are energetically far apart for all values of R, i.e. outside the areas of avoided crossings. In the absence of an electromagnetic field, the spontaneous process may last much longer and even reach seconds. Recently such a process has been measured in the cryogenic ion-beam storage ring DESIREE for small copper and silver clusters of anions\cite{Hansen2017,Anderson2018} and very recently for silver dimer anions\cite{Anderson2020}. During the first direct observation of the spontaneous single-molecule decay of an internally hot diatomic silver anion molecule (Ag$_2^-$), two reaction pathways were discovered. The first path is the fragmentation of the molecular anion into a neutral atom (Ag) and an atomic anion (Ag$^-$), the second is the spontaneous emission of an electron from the molecular anion. The decay process was explained by the Morse potential energy curves generated from the experimental parameter values for neutral and anion sliver dimers. 

Our goal is to theoretically describe the spontaneous decay process of silver anion dimer. The first step is to calculate the potential energy ground states of the anion and neutral silver dimers. Based on the rovibrational levels, we are able to characterise the decay channels and compare our results with the experimental data. We focus on the critical value of the rotational quantum number $J_c$ related to a certain oscillating quantum number v. Having the ground state potential energy curve of an anionic system, we calculate the quantum dynamics (QD) of the dissociation process, which allows us to calculate the widths and lifetimes of high-lying rovibrational levels. The predissociation lifetimes for quasibond states with small widths of levels are treated with a time-independent approach. We also present the non-adiabatic coupling matrix element between chosen initial and final vibrational states of Ag$_2^-$ and Ag$_2$, which allow us to estimate the spontaneous electron emission lifetimes.

Our calculation methods are described in Sec.~\ref{sec:level2}. In Sec.~\ref{sec:level3}, we discuss the results and compare them with the available experimental data. Conclusions are given in Sec.~\ref{sec:level4}.

\section{\label{sec:level2} Computation methods}

\subsection{\label{sec:PEC} Potential energy curves}

We treat the calculations of the interaction between the silver atom and its negative ion analogously to the interaction between two neutral silver atoms. The same electric axial field of the nuclei affects all electrons. The only significant change resulting from the different number of electrons lies in the different configurations of the electronic states. Since the effective electrical action on the valence electron in the anion is relatively weak, some convergence problems can be expected. The adiabatic states with which we are concerned are computed with the Born-Oppenheimer approximation, i.e. as solutions of the following time-independent Schr\"odinger equation  
\begin{eqnarray}\label{Eq1} 
H^{el} \Psi_{i}^{el}(\vec{r};{R}) &=&
E_{i}^{el}(R)\Psi_{i}^{el}(\vec{r};{R}),
\end{eqnarray}
where the internuclear distance $R$ is kept fixed, vector $\vec{r}$
represents all electronic coordinates, $H_{el}$ is the electronic
Hamiltonian of a diatomic system, $\Psi_{i}^{el}(\vec{r};{R})$
describes the $i$-th eigenstate of the Hamiltonian, $E_{i}^{el}(R)$
is the corresponding eigenvalue, i.e. the i-th adiabatic potential.
The Hamiltonian of the system can be written as
\begin{eqnarray}\label{Eq2}
H^{el} &=& H_{A} + H_{B} + V_{AB},
\end{eqnarray}
where $H_{A}$ and $H_{B}$ are the Hamiltonians of the isolated atoms
and $V_{AB}$ is the interaction between them\cite{Wiatr2015, Wiatr2018}. In the current approach, electrons from the 4s, 4p, 4d, and 5s subshells are treated explicitly, while the Ag core, containing the nucleus and 28 electrons occupying inner shells, is represented by the pseudo-potential. The total Hamiltonian in Eq.(1, 2) can be expressed
as
\begin{eqnarray}\label{Eq3}
H^{el} &=& T + V.
\end{eqnarray}
Here $T$ represents the kinetic energy operator of the valence electrons, and $V$ is the interaction operator between valence electrons and Ag cores. The latter can be written in the following form:
\begin{eqnarray}\label{Eq4}
V &=& V^{A} + V^{B} + \sum_{j>i=1}^{N} \frac{1}{r_{ij}}
+ V_{cc}.
\end{eqnarray}
Here $V^{A}$ describes Coulomb and exchange interaction as well as
the Pauli repulsion between the valence electrons and the Ag
core. We use the following semi-local energy-consistent
pseudopotentials\cite{Figgen2005}:
\begin{eqnarray}\label{Eq5}
  V^{A} &=& \sum_{i=1}^{N}\bigg{(}-\frac{Q_{A}}{r_{A i}} +
  \sum_{l, k}B^{A}_{l, k}\exp(-\beta^{A}_{l, k}r_{A i}^{2})P^{A}_{l}\bigg{)},
\end{eqnarray}
where $Q_{A} = 19$ denotes the net charge of the Ag-core,
$P^{A}_{l}$ is the projection operator onto the Hilbert subspace of
angular symmetry $l$ with respect to the Ag$^{19+}$-core and the number of the valence electrons is $N=19$ or $20$ for neutral or anion molecules, respectively. The parameters $B^{A}_{l, k}$
and $\beta^{A}_{l, k}$ define the semi-local energy-consistent
pseudo-potentials\cite{Figgen2005}. The second term in equation (\ref{Eq4}) is fully analogous to the first but concerns the second Ag-core. The third term stands for the repulsion
between the valence electrons, whereas the last term describes the
interaction between Ag cores.

The core electrons of the Ag atom are represented by the energy-consistent pseudo-potential\cite{Figgen2005} which  was formed from basis set converged towards the complete basis set limit\cite{Peterson2005}. Note that the spin-orbit (SO) effect makes a very small contribution to the ground energies of our systems, as it only increases the depth of the well by only 24.5 cm$^{-1}$ \cite{Peterson2005}, so we do not include it in our calculations.

The first goal of our calculations is to recover the electron affinity (EA) of silver, knowing that the experimental value given by Bilodeau\cite{Bilodeau_1998} is 10521 cm$^{-1}$. We perform CCSD(T) calculations with Peterson's aug-cc-pV5Z-PP basis set\cite{Peterson2005}, but the obtained result equals 10215 cm$^{-1}$ is 306 cm$^{-1}$ below the experimental value. Then we decide to take the core-valence correlation into account and switch cc-pwCV5Z-PP basis set, augmented with the aug-cc-pV5Z-PP exponents. The obtained values of 10608 cm$^{-1}$ is much closer to the experimental one but overestimates it by 87 cm$^{-1}$. Interestingly, the aug-cc-pV5Z-PP basis set with core-valence correlation turned on leads to values almost the same as for the aug-cc-pwCV5Z-PP basis sets.
    
The problem may lie in the objective function (OF) used in the optimisation of Peterson's basis set, which was set as average singles-doubles configuration interaction (CISD) energy of $ns^2(n-1)d^9$ and $ns^1(n-1)d^{10}$ states of a neutral atom. In our case, we are only interested in the ground states of Ag and Ag$^-$, so we decide to optimize our basis set using the sum of the lowest energy of the neutral state Ag and its anion with the following objective function:
\begin{eqnarray}
\label{dk.of}
OF&=&E_{FC_{CCSD},\,Ag}+E_{FC_{CCSD},\,Ag^-}\nonumber\\
  &+&0.5(E_{CV_{CCSD},\,Ag}+E_{CV_{CCSD},\,Ag^-})\nonumber\\
  &+&0.01(E_{CC_{CCSD},\,Ag}+E_{CC_{CCSD},\,Ag^-}),
\end {eqnarray}
where $E_{X,\,Y}$ is the electronic energy component $X$: $FC_{CCSD}$-frozen-core CCSD energy, $CV_{CCSD}$ core-valence corelation CCSD energy and $CC_{CCSD}$ core-core corelation CCSD energy of the system $Y=\{Ag,\,Ag^-\}$. Such an objective function may be seen as a more convenient alternative to the three-step method presented by Puzzarini\cite{Peterson2005}, in which the HF energy and weighted correlation energy components were optimized separately. We keep the number of exponents of the augmented cc-pwCV5Z-PP basis. Instead of directly optimizing the Gaussian exponents, according to Peterson\cite{Peterson2003}, we expand them into orthonormal Legendre polynomials, up to six terms, and optimize the expansion coefficients. After finalizing the optimization step,  we contract the $s$, $p$ and $d$ exponents using atomic natural orbital (ANO) coefficients. In addition, we obtain uncontracted mid-bond (MB) 8s6p4d2f exponents optimized on the sum of the frozen-core CISD energy of Ag$_2$ (at the experimental R$_e$=2.5303 \AA)\cite{Laher2011} and Ag$_2^-$ (at the experimental R$_e$=2.604 \AA)\cite{Ho1990} calculated with augmented cc-pwCVQZ-PP basis set.

To calculate adiabatic potential energy curves of Ag$_2$ and Ag$_2^-$ diatomic molecules we use the  multiconfigurational self-consistent field/complete active space self-consistent field (MCSCF/CASSCF) method followed by the multi-reference configuration interaction (MRCI) method and single reference coupled cluster method with singles and doubles, and perturbative triples (CCSD(T)). The corresponding active space includes the molecular counterparts of the 4s, 4p, 4d, and 5s orbitals of the Ag atoms. The interaction energies are corrected for the basis set superposition error (BSSE) by counterpoise (CP) correction. For comparison, we also perform calculations with the original basis sets of Peterson and Puzzarini \cite{Peterson2005} and we assign the obtained potential curves as PP. All calculations are performed by means of the MOLPRO program package \cite{MOLPRO}. Using these computational methods, we obtain adiabatic potential energy curves for the ground state $^2\Sigma_u^+$ of Ag$_2^-$ and the ground state $^1\Sigma_g^+$ of Ag$_2$. The new sPYtroscopy\cite{Jasik2020} program is used to calculate rovibrational levels, spectroscopic parameters and the non-adiabatic coupling matrix element between the selected initial and final vibrational states of Ag$_2^-$ and Ag$_2$.

\subsection{\label{sec:level22b} Non-adiabatic coupling elements}

Before the ejection of an electron, the molecular anion Ag$_2^-$ is in the 
electronic state $1^2\Sigma_u^+$ with the electronic wave function $\Psi_i^{(N+1)}(\vec{r}_1,\cdots,\vec{r}_{N+1};R)$.
After the electron is ejected, the wave function is given by
\begin{eqnarray*}
\Psi_f(\vec{r}_1,\cdots,\vec{r}_{N+1};R;E_{\rm cont}) &=& {\cal A} \left( \Psi_f^{N}(\vec{r}_1,\cdots,\vec{r}_N;R) \right.\\
&& \left. \cdot \phi_{\rm cont}(\vec{r}_{N+1};R;E_{\rm cont}) \right).
\end{eqnarray*} 
Here, $\cal A$ is the antisymmetrisation operator and $E_{\rm cont}$ is the 
energy of the ejected electron. $\Psi_f^{N}(\vec{r};R)$ is the wave function of the 
the electronic ground state $1^1\Sigma_g^+$ of the neutral molecule Ag$_2$ and
$\phi_{\rm cont}(\vec{r}_{N+1};R;E_{\rm cont})$ is the continuum wave function of the ejected electron. 

The non-adiabatic coupling element 
between two electronic states is given by
\begin{eqnarray}\label{Eq:electronic coupling}
\Lambda_{fi}(R,E_{\rm cont}) &=& \int 
\Psi_f^*(\vec{r}_1, \cdots, \vec{r}_{N+1};R;E_{\rm cont}) \nonumber \\
&& \frac{\partial}{\partial R} 
\Psi_i(\vec{r}_1, \cdots, \vec{r}_{N+1};R)
d\vec{r}_1 \cdots d\vec{r}_{N+1},
\end{eqnarray}
where integration includes all electrons. The coupling elements depend on the internuclear distance $R$ and the energy $E_{\rm cont}$ of the ejected electron. Here we neglect the terms with second derivatives. The non-adiabatic coupling matrix element between initial and final vibrational states is calculated as  
\begin{equation}\label{Eq:vibronic coupling}
\Lambda_{v'f,vi}(E_{cont}) = \frac{\hbar^2}{\mu_{Ag_2}}\int \chi_{v'f}^*(R) \Lambda_{fi}(R,E_{\rm cont})
 \frac{\partial}{\partial R} \chi_{vi}(R) dR,
\end{equation}
where $\mu_{Ag_2}$ is the reduced mass of the Ag$_2$ molecule. $\chi_{vi}$ is the initial vibrational wave function of the anion and $\chi_{v'f}$ is the final vibrational wave function of the neutral molecule.

The probability of the electron autodetachment per unit time is proportional to the square of the element of the  non-adiabatic vibrational coupling matrix\cite{Douguet2015} and can be expressed as follows
\begin{equation}\label{Eq:ad level width}
\Gamma_{ad} \sim \frac{2\pi}{\hbar} |\Lambda_{v'f,vi}|^2\, .
\end{equation}
The lifetimes of the spontaneous electron emission $\tau$ are then estimated as the inverse of $\Gamma_{ad} = \hbar/\tau$.

The essential part of the above approach is the computation of the non-adiabatic coupling matrix elements $\Lambda_{fi}$ between the ground states of the molecular anion and the neutral molecule. Neutral silver dimer in the electronic ground state has a closed-shell electron configuration. The ab initio wave function of the electronic ground state is dominated by the Hartee-Fock determinant. Therefore, in the computation of the electronic coupling elements, the many-electron wave functions are approximated by single Slater determinants with Kohn-Sham orbitals. The wave function of the anion is constructed by adding an extra electron into 
the lowest unoccupied orbital $\phi_{\rm bound}$ of the neutral molecule and 
assuming that the occupied orbitals do not change.
The non-adiabatic coupling element simplifies to
\begin{equation} \label{Eq:coupling}
\Lambda_{fi}(R,E_{\rm cont}) = \int 
\phi_{\rm cont}^*(\vec{r},R;E_{\rm cont}) 
\frac{\partial}{\partial R} 
\phi_{\rm bound}(\vec{r},R)
d\vec{r}.
\end{equation}
Here the integration is over the coordinates of the ejected electron.

The continuum wave function $\phi_{\rm cont}(\vec{r},R;E_{\rm cont})$ of the ejected electron is computed in a quantum scattering calculation using the single centre expansion (SCE) method \cite{sce} and Kohn-Sham density functional theory.
The scattering potential that describes the interactions between the ejected electron and the neutral 
molecule is given by
\begin{equation}
v_{tot} = v_{st} + v_{cp} + v_{ex},
\end{equation}
where $v_{st}$ is the electrostatic potential,
$v_{cp}$ is the correlation-polarisation potential, 
and $v_{ex}$ is the exchange potential.
All potentials are functions of the electron density of the neutral molecule.
The correlation-polarisation potential is approximated using the 
generalized gradient approximation of Lee, Yang and Parr (LYP) \cite{b3lypLYP}.
For large distances, the potential is switched to the asymptotic form 
\begin{equation}
v_{cp} \rightarrow  - \frac{\alpha}{r^4}, 
\end{equation}
where $\alpha$ is the polarisability of the neutral molecule.
The switching between the LYP and the long-range potential is done 
by a switching function around the outermost crossing point $r_c$ of the two potentials.
The exchange potential is approximated by the modified semi-classical exchange potential\cite{vmsce}.
For large distances ($r \ge r_c$) from the target the exchange potential is 
represented by the form similar to the one suggested by Silkowski and Pachucki\cite{vexasym}
\begin{equation}
v_{ex} \rightarrow -b r^{5/2} e^{-ar} \;,
\end{equation}
where the parameters $a$ and $b$ are determined by fitting to the radial points around $r_c$.

All calculations have been performed for a number of nuclear geometries.
In each geometry, a single centre expansion of the bound orbital $\phi_{\rm bound}(\vec{r},R)$
is done with SCELib4 of Sanna {\it et al.} \cite{scelib4}. 
The expansion is largely dominated, with a coefficient greater than 0.9, by the $p$-wave 
that is aligned along the molecular axis. 
For all further calculations only the p-wave is considered,
$\phi_{\rm bound}(\vec{r},R)$ is renormalised,
and
$\frac{\partial}{\partial R} 
\phi_{\rm bound}(\vec{r},R)$
is computed by numerical differentiation using central-difference derivatives. 

For each nuclear geometry the electron density is computed with density functional theory. These computations are done with the program package Gaussian \cite{gaussian}  
using the  B3LYP (Becke\cite{b3lypBecke}, three-parameter, Lee-Yang-Parr\cite{b3lypLYP}) 
exchange-correlation functional,
the effective core potentials of Hay and Wadt\cite{ecphay1,ecphay2}, 
and the double-zeta basis set of Dunning and Hay\cite{dunning-hay}.
The single centre expansion of the electron density is done with the 
program package SCELib4\cite{scelib4}.

The electron scattering calculations to compute 
$\phi_{\rm cont}(\vec{r},R;E_{\rm cont})$ 
are done 
with the program package Bumblebee \cite{bumblebee}. 
The program solves the scattering equations in the presence
of the potential $v_{tot}$ by a hybrid propagator.
For short ranges (up to about 10 \AA) R-matrix propagation \cite{light78} is used.
For large ranges, the variable phase approach \cite{calogero} is used and the equations are integrated to the distance where the phase shifts converge using the 
embedded Runge-Kutta Prince-Dormand (8,9) method\cite{dormandprince}. In the asymptotic region the normalisation of the wave function is done and the boundary conditions are propagated inwards. The method of Kulander and Light\cite{light80} is used to construct 
the wave function from the R-matrices\cite{franz2011}.
Whenever possible the program uses the algorithms implemented in the GNU scientific library (GSL) \cite{gsl}. Only the s-wave and the p-wave (aligned along the molecular axis) are considered. For the evaluation of the integral in Eq. (\ref{Eq:coupling}) the integrand is represented
as Akima-spline\cite{akima}. A spline is a piecewise polynomial, which is integrated analytically, as discussed in Chapter 9 in Ueberhuber\cite{ueberhuber}.

\subsection{\label{sec:level23} Quantum dynamics}

The time-dependent approach which is mathematically equivalent to the time-independent one can be regarded as a complementary tool giving other insight into physical processes\cite{Jasik2017, Jasik2018}. In our case, we use the time-dependent method to investigate the rotational predissociation of highly excited rovibrational states of the Ag$_2^-$, which is one of the competitive reactions in the problem under consideration. We start our consideration from the time-dependent Schr\"{o}dinger equation written in the following form
\begin{equation}\label{TDSE}
\imath \hbar \frac{\partial}{\partial t} \Phi(R,t) = H_{J}^{nuc}
\Phi(R,t),
\end{equation}
where $\Phi (R, t)$ is the time-dependent wave packet moving along the effective potential energy curve, and $H_ {J} ^ {nuc}$ is the normal nuclear Hamiltonian.

By definition the wavepacket is a coherent superposition of
stationary states (e.g. Tannor
\cite{Tannor2007}) which may be represented in the following form consisting of two contributions from the discrete and  continuous parts of the spectrum
\begin{equation*}
\Phi({R}; t) = \sum_{v,J} c_{v,J} \, \Psi^{nuc}_{v,J}(R) \,  \mathrm{e}^{-\imath E(v,J)t / \hbar} \; +
\end{equation*}
\vspace{-0.55cm}
\begin{equation}\label{pakiet falowy}
 + \int\, c_{J}(E) \, \Psi_{E ,J}(R)\, \mathrm{e}^{-\imath Et / \hbar} \, dE, 
\end{equation}
where
\begin{equation*}
c_{v,J} \ = \int_0^{\infty} \,
\Psi_{\nu,J}^{nuc}(R)^{\!*} \, \Phi(R;0) \, dR
\end{equation*}
and
\begin{equation*}
c_{ J}(E) \ = \int_0^{\infty} \, \Psi_{E,J}(R)^{\!*} \, \Phi(R;0) \, dR
\end{equation*}
are the energy-dependent coefficients, squares of these coefficients form the spectral distribution of $\Phi$ normalized to 1, $\mathrm{e}^{-\imath E(v,J)t / \hbar}$ and 
$\mathrm{e}^{-\imath Et / \hbar}$ are the time evolution factors, 
$\Psi^{nuc}_{v,J}(R)$ and $\Psi_{E ,J}(R)$ are eigenfunctions of ${H}_ J^{nuc}(R)$. The wavepacket $\Phi(R; t)$ is a solution of Eq.\ref{TDSE} and
its initial shape at $t = 0$  is taken as a Gaussian function of
arbitrary half-width placed on the effective potential energy curve. The
wavepacket moves away from its starting location due to the Newtonian force
$-dU_J/dR$. This process is described by the time-dependent
autocorrelation function
\begin{equation}\label{funkcja autokorelacji}
S(t) =  \int \Phi({R}; t=0)^{\!*} \,  \, \Phi({R}; t) \, dR.
\end{equation}
In our case, the autocorrelation function describes evolution of the initial nuclear wavepacket in the ground electronic state of the molecular anion Ag$_2^-$. The time-dependent  population  in the range till $R_{max}$ for the particular state labeled  by $J$, in accordance with the effective potential energy $U_J$, is calculated as
\begin{equation}\label{population_J}
P(t) =  \int_0^{R_{max}} | \Phi({R}; t) |^2\, dR.
\end{equation}
We determine the spectrum by the inverse Fourier transform of
$S(t)$~\cite{Schinke1993,Bilingsley1995} as follows
\begin{equation}\label{tacs}
\sigma(E(\nu,J)) 
= \int_{-\infty}^{\infty} \mathrm{e}^{\imath E(\nu, J) t / \hbar} \, S(t) \, dt.
\end{equation}

In our calculations, the above integral is estimated in the range <0,T> using Fast Fourier Transform (FFT) procedures~\cite{2005Frigo}.

The propagation time is assumed to be 250 ps, which is sufficient to estimate the integral in Eq. 17. In Eq. 16, we set $R_{max}$ to be equal to 100 a$_0$. 
There are 2$^{14}$ points in the integration grid. To avoid the diffraction between the outgoing and the incoming waves as a result to bouncing off the boundary at $R_{max}$, the negative imaginary potential is placed at 60 a$_0$. This potential smoothly absorbs the wavepacket near the boundary. All quantum-dynamic computations are performed using parts of the YADE platform\cite{yade:manual2,yade:Kozicki2008,yade:Kozicki2009,yade:YadePubs}.
  
\section{\label{sec:level3} Results and discussion}
\begin{table*}[th!]
\caption{\label{tab:table1}Spectroscopic parameters R$_e$ [\AA], D$_e$, D$_0$ $\omega_e$, B$_e$, and EA [cm$^{-1}$] of the 1$^1\Sigma_g^+$ state of Ag$_2$. PP represents the original aV5Z/awCV5Z basis set of Peterson and Puzzarini\cite{Peterson2005}, CP represents the counterpoise correction, MB represents the mid-bonds, and MB+CP represents the mid-bonds with counterpoise correction.}
\begin{ruledtabular}
\begin{tabular}{lcccccc}
Author & R$_e$ & D$_e$ & D$_0$ & $\omega_e$ & B$_e$ & EA Ag$_2$ \\
\hline
present, MRCI & 2.570 & 11165 & 11078 & 175.6 & 0.04765 & 8377 \\
\hline
present, CCSD(T)             & 2.523 & 13743 & 13645 & 195.8 & 0.04946 & 8094 \\
present, CCSD(T) CP             & 2.523 & 13695 & 13597 & 195.6 & 0.04941 & 8065 \\
present, CCSD(T) MB             & 2.517 & 14091 & 13992 & 197.2 & 0.04962 & 8088 \\
present, CCSD(T) MB+CP          & 2.523 & 13825 & 13727 & 195.9 & 0.04944 & 8053 \\
\hline
present PP, CCSD(T)         & 2.523 & 13825 & 13727 & 196.1 & 0.04947 & 8159 \\
present PP, CCSD(T) CP      & 2.523 & 13750 & 13653 & 195.4 & 0.04938 & 8163 \\
present PP, CCSD(T) MB      & 2.517 & 14273 & 14174 & 198.5 & 0.04972 & 8120 \\
present PP, CCSD(T) MB+CP   & 2.523 & 13844 & 13746 & 196.0 & 0.04945 & 8145 \\
\end{tabular}
\end{ruledtabular}
\end{table*}
\begin{table*}[th!]
\caption{\label{tab:table2}Spectroscopic parameters R$_e$ [\AA], D$_e$, D$_0$, $\omega_e$, and B$_e$ [cm$^{-1}$] of the 1$^2\Sigma_u^+$ state of Ag$_2^-$. PP represents the original aV5Z/awCV5Z basis set of Peterson and Puzzarini\cite{Peterson2005}, CP represents the counterpoise correction, MB represents the mid-bonds, and MB+CP represents the mid-bonds with counterpoise correction.}
\begin{ruledtabular}
\begin{tabular}{lccccc}
Author & R$_e$ & D$_e$ & D$_0$ & $\omega_e$ & B$_e$ \\
\hline
present, MRCI           & 2.685 & 10407 & 10339 & 136.7 & 0.04372 \\
\hline
present, CCSD(T)                & 2.642 & 11288 & 11215 & 146.2 & 0.04493 \\
present, CCSD(T) CP             & 2.649 & 11211 & 11138 & 145.9 & 0.04485 \\
present, CCSD(T) MB             & 2.636 & 11631 & 11557 & 148.6 & 0.04516 \\
present, CCSD(T) MB+CP          & 2.649 & 11329 & 11256 & 146.3 & 0.04490 \\
\hline
present PP, CCSD(T)         & 2.642 & 11351 & 11278 & 146.8 & 0.04495 \\
present PP, CCSD(T) CP      & 2.649 & 11280 & 11207 & 146.1 & 0.04483 \\
present PP, CCSD(T) MB      & 2.636 & 11761 & 11686 & 150.2 & 0.04531 \\
present PP, CCSD(T) MB+CP   & 2.642 & 11356 & 11283 & 146.5 & 0.04492 \\
\end{tabular}
\end{ruledtabular}
\end{table*}
\begin{table*}[ht!]
\caption{\label{tab:table3}The comparison of present spectroscopic parameters R$_e$ [\AA], D$_e$, D$_0$ $\omega_e$, B$_e$, and EA [cm$^{-1}$] of the 1$^1\Sigma_g^+$ state of Ag$_2$ with other experimental and theoretical results.}
\begin{ruledtabular}
\begin{tabular}{lcccccc}
Author & R$_e$ & D$_e$ & D$_0$ & $\omega_e$ & B$_e$ &  EA Ag$_2$ \\
\hline
present, CCSD(T) CP & 2.523 & 13695 & 13597 & 195.6 & 0.04941 & 8065 \\
Laher (exp.)\cite{Laher2011} & 2.530 & & 13308 & 192.4 & 0.04881 & \\
Morse (exp.)\cite{Morse1986} & 2.480 & & 13308 $\pm$ 242 & 192.4 & 0.05121 \\
Peterson and Puzzarini (theory) \cite{Peterson2005} aV5Z/awCV5Z & 2.524 & 13714 &  & 196.0 &  &  \\
Peterson and Puzzarini (theory) \cite{Peterson2005} CBS & 2.523 & 13766 &  & 196.2 &  &  \\
Hay et al. (theory)\cite{Hay1985} & 2.617 & 10703 &  & 172.0 &  & \\
Zhang et al. (theory)\cite{Zhang1993} & 2.644 &  &  & 160.0 &  & \\
Stoll et al. (theory)\cite{Stoll_1984} & 2.688 & 13147 &  & 142.0 &  &  \\
Bonacic-Koutecky et al. (theory)\cite{Bonacic1994} & &  &  &  &  & 8630 \\
Ho et al. (exp.)\cite{Ho1990} &  &  &  &  &  & 8251 $\pm$ 56.5 \\
\end{tabular}
\end{ruledtabular}
\end{table*}
In Tab.~\ref{tab:table1} and Tab.~\ref{tab:table2}, we present all our spectroscopic parameters for Ag$_2$ and Ag$_2^-$ molecules, respectively. All calculations are performed using our optimised 5Z-type basis set and the original basis set aV5Z/awCV5Z provided by Peterson and Puzzarini\cite{Peterson2005} (PP) and the different computational approaches described in Section~\ref{sec:PEC}.

An in-depth analysis of the calculated spectroscopic parameters and potential energy curves carried out in relation to all available experimental results (see Tab.~\ref{tab:table3} and Tab.~\ref{tab:table4}) allows us to conclude that for both systems the best agreement is obtained using the optimised basis set and coupled-cluster (CC) method with counterpoise correction.
\begin{table*}[ht!]
\caption{\label{tab:table4}The comparison of spectroscopic parameters R$_e$ [\AA], D$_e$, D$_0$, $\omega_e$, and B$_e$ [cm$^{-1}$] of the 1$^2\Sigma_u^+$ state of Ag$_2^-$ with other experimental and theoretical results.}
\begin{ruledtabular}
\begin{tabular}{lccccc}
Author & R$_e$ & D$_e$ & D$_0$ & $\omega_e$ & B$_e$ \\
\hline
present, CCSD(T) CP & 2.649 & 11211 & 11138 & 145.9 & 0.04485 \\
Bonacic-Koutecky et al. (theory)\cite{Bonacic1994} & 2.780 & &  & & \\
Spasov et al. (theory)\cite{Spasov1999} & & &  & & 0.04220\\
Ho et al. (exp.)\cite{Ho1990} & 2.604 $\pm$ 0.007 & & 11050 $\pm$ 1291 & 145 $\pm$ 10 & \\
\end{tabular}
\end{ruledtabular}
\end{table*}
For the selected computational approach, the estimated root mean squared error (RMSE) and mean absolute error (MAE) are the smallest with respect to the experimental data and are respectively 126 and 71. The same errors calculated for potentials obtained by using PP basis set and CC method with CP correction are larger reaching values equal to 138 and 74. Other combinations of basis sets and methods lead to much larger errors.
\begin{figure*}[ht!]
    \centering
    \includegraphics[width=\textwidth]{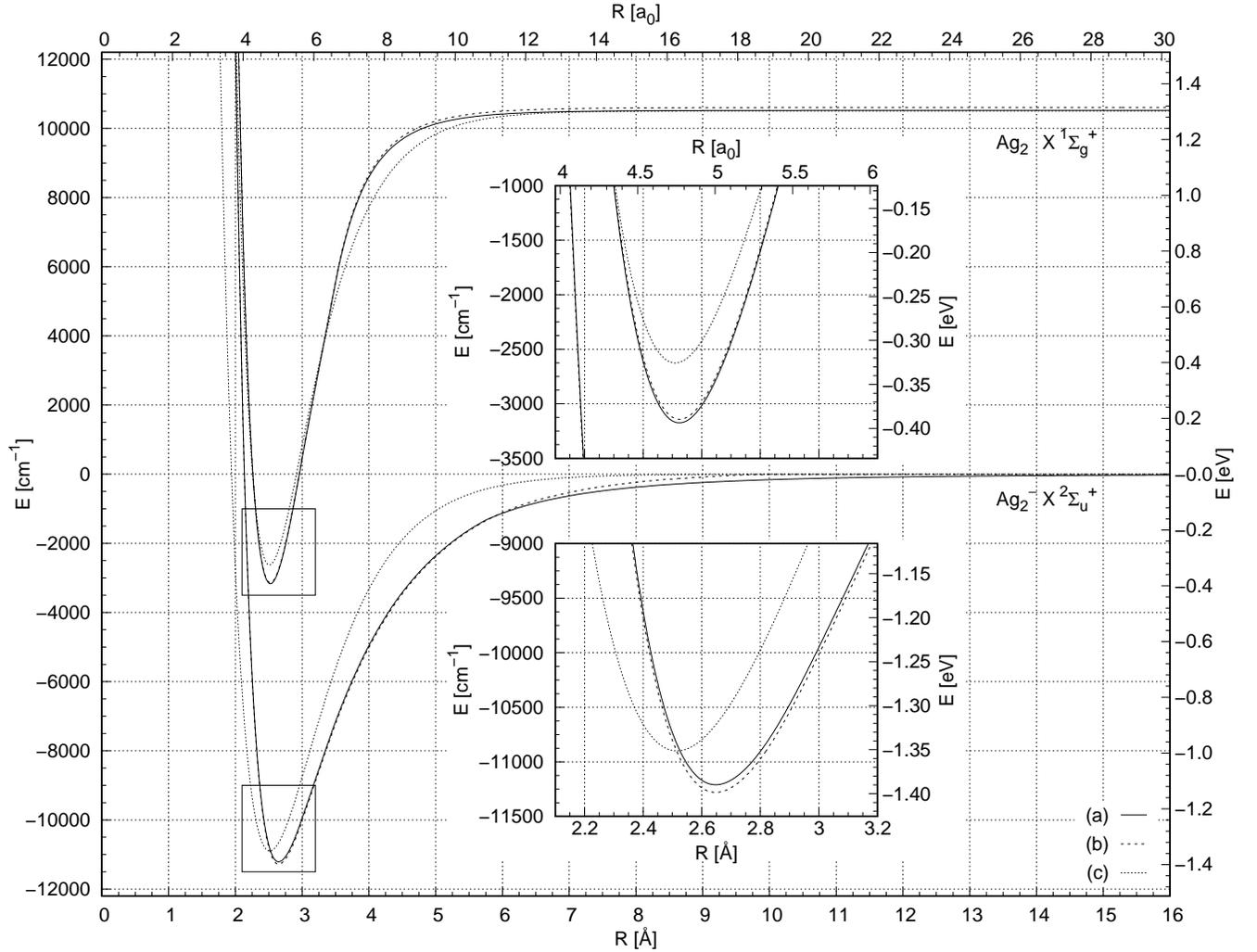}
    \caption{\label{fig:BO pot} Potential energy curves for the Ag$_2^-$ and Ag$_2$ ground states: (a) current work; (b) current work PP; (c) Morse potentials generated by experimental and theoretical parameters\cite{Anderson2020}.}
\end{figure*}

In Fig. \ref{fig:BO pot} we show our potential energy curves calculated with the use of selected approaches (optimised and PP basis sets + CC with CP) for the ground states of both systems. A visual comparison can be made with the Morse potential curves\cite{Anderson2020} plotted for Ag$_2^-$ with the experimental parameters\cite{Ho1990} and for Ag$_2$ with the theoretical parameters\cite{Hay1985}. In the case of Ag$_2^-$, the divergence between the two curves is quite pronounced, especially for the attractive part of the potential and smaller for the repulsive part. In the case of Ag$_2$, the differences in the positions of the potential wells between the parameterized Morse potential\cite{Anderson2020} and our theoretical results remain within a reasonable range. On the other hand, for Ag$_2^-$, the analogous difference is much greater, but it should not seriously affect the interpretation of the experimental data.

For Ag$_2$ spectroscopic parameters (see Tab.~\ref{tab:table3}), our results show better agreement with more recent Laher experimental data\cite{Laher2011} than those reported by Morse\cite{Morse1986}. As for Ag$_2^-$ (see Tab.~\ref{tab:table4}), the agreement between our results and the only available experimental data of Ho et al.\cite{Ho1990} is quite satisfactory.

The electron affinity (EA) value plays a key role. In fact, the spontaneous emission of electrons  is allowed because EA of Ag$_2$ is less than the dissociation energy of Ag$_2^-$. Then in Tab.~\ref{tab:table3} and \ref{tab:table5_EA}, we show our calculated molecular and atomic EA. Although  the atomic values are fully in line with the experimental data, the molecular values remain around 200 cm$^{-1}$ below the experimental data.
\begin{table}[ht!]
\caption{\label{tab:table5_EA}Electron affinity EA [cm$^{-1}$] of the Ag atom.}
\begin{ruledtabular}
\begin{tabular}{lc}
Author & EA Ag \\
\hline
Bilodeau et al. (exp.)\cite{Bilodeau_1998} & 10521 \\
present, CCSD(T) CP        & 10524 \\
present PP, CCSD(T) CP & 10608 \\
present, MRCI   & 9116 \\
Peterson and Puzzarini \cite{Peterson2005} CBS & 10598 \\
Peterson and Puzzarini \cite{Peterson2005} aV5Z/awCV5Z & 10573 \\
Stoll et al. \cite{Stoll_1984} & 9033 \\
\end{tabular}
\end{ruledtabular}
\end{table}
 
\begin{figure}[ht!]
    \centering
    \includegraphics[width=\columnwidth]{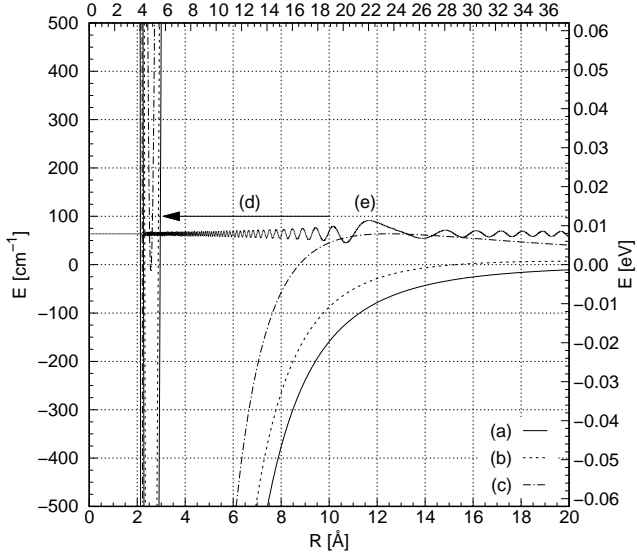}
    \caption{\label{fig:BO1 pot} Effective potential energy curves of Ag$_2^-$ and Ag$_2$ ground states; (a) $J$=0; (b) $J$=150; (c) $J$=254; (d) spontaneous electron emission is no longer possible for $J\geq$254; (e) vibrational wave function for $v$=155, $J$=254.}
\end{figure}

\begin{figure}[ht!]
    \centering
    \includegraphics[width=\columnwidth]{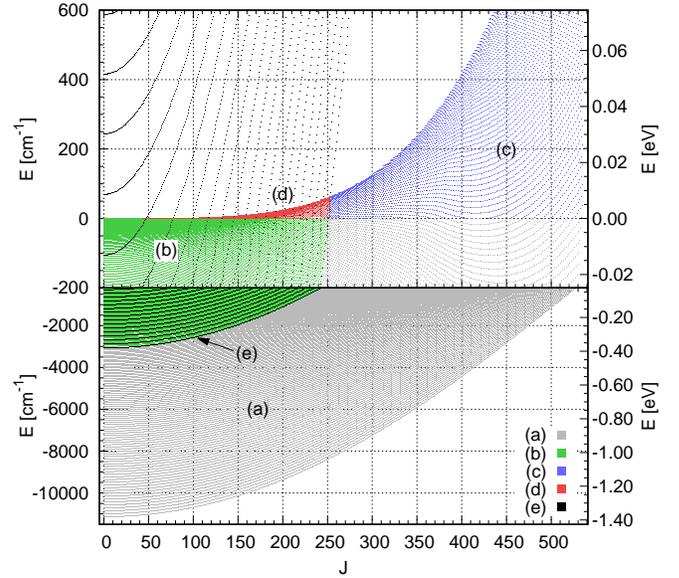}
    \caption{\label{fig:Energy-J} A map of rovibrational levels of Ag$_2^-$ and Ag$_2^{}$:
    (a) no decay possible, the gray points are rovibrational levels of Ag$_2^{-}$, but only those lying below the lowest level of Ag$_2^{}$;
    (b) spontaneous electron emission only, the green points are rovibrational levels of Ag$_2^{-}$, but only those lying above the lowest levels of Ag$_2^{}$ and below zero;
    (c) fragmentation only, the blue points are quasi-bound rovibrational levels of Ag$_2^{-}$, no mixing with rovibrational levels of Ag$_2^{}$;
    (d) both decay modes possible, the red points are quasi-bound rovibrational levels of Ag$_2^{-}$, but only those mixing with the rovibrational levels of Ag$_2^{}$;
    (e) the black points are rovibrational levels of Ag$_2^{}$.
    }
\end{figure}

The analysis of the effective potential energy curves along with the increase of the rotational quantum number $J$ (Fig.~\ref{fig:BO1 pot}) shows that from $J = 254$, spontaneous emission of electrons from Ag$_2^-$ is not possible.
Then we define $J_c$ = 253 as the critical value, i.e. the highest  $J$ value for which spontaneous emission is allowed. In this case, only the lowest vibration level of the neutral system lies below the highest vibration levels of the anionic system. Our map of the rovibrational levels of Ag$_2^-$ and Ag$_2$ molecules (Fig.~\ref{fig:Energy-J}) shows the regions defined by the rotational quantum numbers $J$ and energies. A much more accurate division of the energy scale above -200 cm$^{-1}$ allows for better recognition of transitions between different decay paths. What is not visible everywhere, each line has a discrete structure. Each part of these lines corresponds to an energy of a given rovibrational level defined by a pair of quantum numbers $(v, J)$.

\begin{figure*}[ht!]
    \centering
    \includegraphics[width=\textwidth]{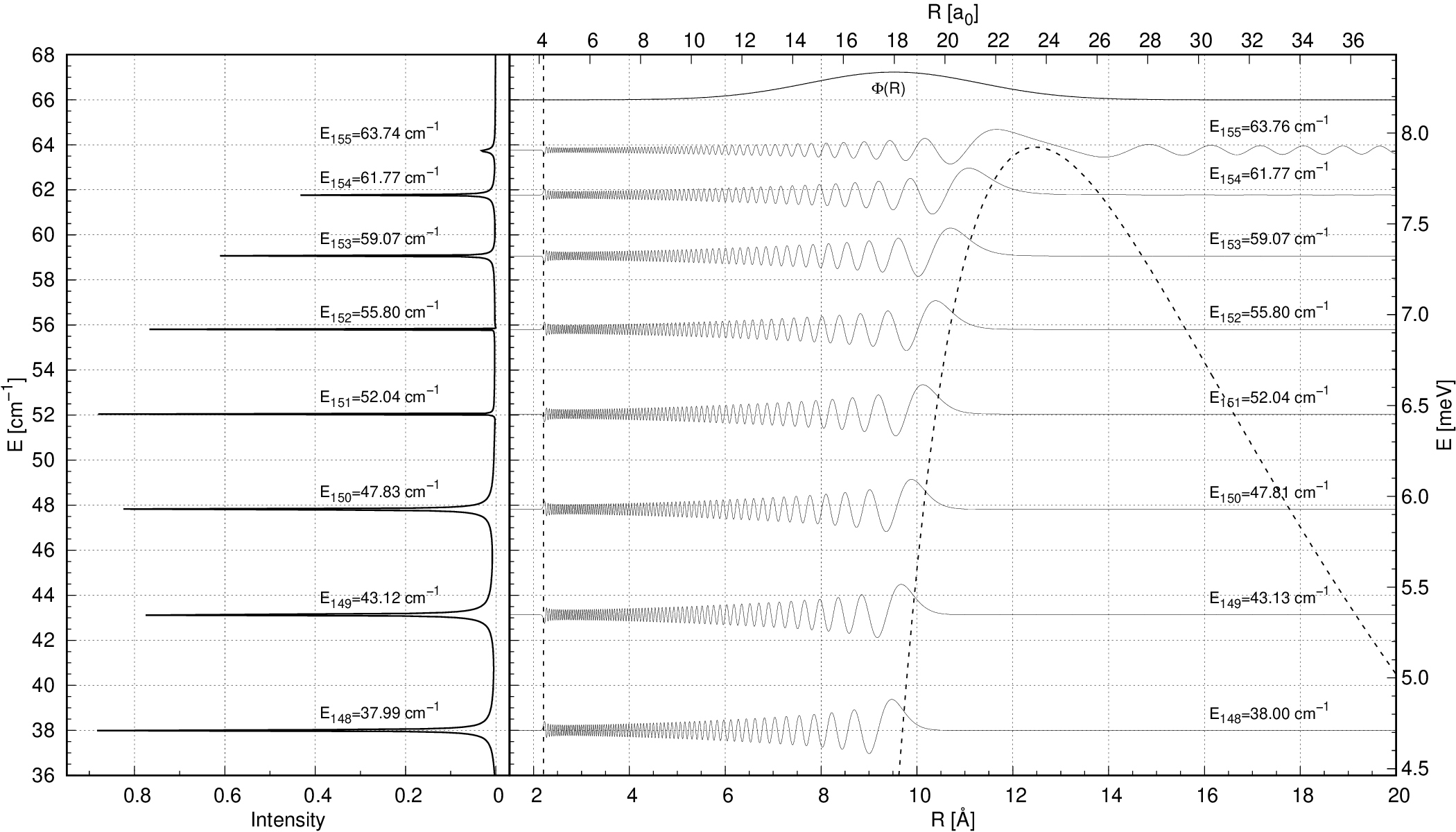}
    \caption{\label{fig:gauss_packet} (left panel) Vibrational energy levels for the effective potential with $J_c+1=254$ obtained by autocorrelation function (Eq.~\ref{tacs}) calculated from quantum dynamics of the Gaussian packet $\Phi({R};t=0)$ (shown on the top of the right panel) using the YADE software~\cite{yade:manual2,yade:Kozicki2008,yade:Kozicki2009,yade:YadePubs,Jasik2018} are compared with energy levels and vibrational wave functions derived from the sPYtroscopy program\cite{Jasik2020} (right panel).}
\end{figure*}

The spectrum of the effective potential energy curve for $J_c + 1$ obtained using quantum dynamics in the YADE framework\cite{yade:manual2,yade:Kozicki2008,yade:Kozicki2009,yade:YadePubs} (the autocorrelation function in Eq.~\ref{tacs}) from the Gaussian wavepacket $\Phi({R};t=0)$ is presented in Fig.~\ref{fig:gauss_packet} and compared with the energies of the vibrational levels obtained from the sPYtroscopy program\cite{Jasik2020}. QD calculations are very accurate for quasibond states with level widths larger than approximately 10$^{-4}$ cm$^{-1}$ as opposed to time-independent methods implemented in Level\cite{LeRoy2017} or sPYtroscopy\cite{Jasik2020}. We calculate all the level widths and predissociation lifetimes for the rotational quantum number J from 0 to J$_c$+1 = 254 using QD for highly excited rovibrational levels of Ag$_2^-$ and time-independent methods for all other rovibrational levels. Based on these calculations, and assuming that the maximal lifetime of the fragmentation reaction is less than 10 seconds, which corresponds to the duration of the experiment reported by Anderson \emph{et al.}\cite{Anderson2020}, we estimate the mean value of the predissociation lifetime to be approximately 224~ms.

For quasibond states with large level widths, the lifetimes of highly-lying rovibrational levels are calculated from the vibrational wave functions using the recently described method\cite{Jasik2018}. The obtained exemplary lifetimes (by fitting the population to the exponential decay $e^{-t/\tau}$)~\cite{Jasik2018} are $\tau_{(J = 254, v = 154)}=204~{\rm{ns}}$ and $\tau_{(J = 254, v = 155)}=57~{\rm{ps}}$, where the latter one corresponds to the highest vibrational level. Their corresponding level widths (i.e. the full width at half maximum, FWHM) are $\Gamma_{(J = 254, v = 154)}^{YADE}=0.26\times{10^{-4}}~{\rm{cm}}^{-1}$ and $\Gamma_{(J = 254, v = 155)}^{YADE}=0.093~{\rm{cm}}^{-1}$. The respective values obtained from the LEVEL program\cite{LeRoy2017} using time-independent treatment are $\Gamma_{(J = 254, v = 154)}^{LEVEL}=0.15\times{10^{-4}}~{\rm{cm}}^{-1}$ and $\Gamma_{(J = 254, v = 155)}^{LEVEL}=0.072~{\rm{cm}}^{-1}$. The above analysis leads us to the statement that the fragmentation reaction (Ag$_2^-$ $\rightarrow$ Ag + Ag$^-$) for the highly-lying rovibrational levels is very fast. But for quasibond states with smaller level widths, it reaches milliseconds or even a few seconds.
\begin{figure}[ht!]
    \centering
    \includegraphics[width=\columnwidth]{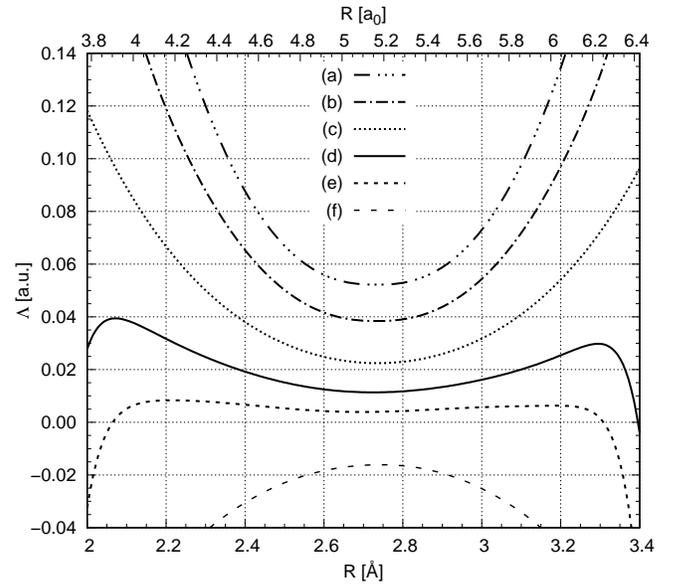}
    \caption{\label{fig:coupling_J0} Electronic non-adiabatic coupling elements for chosen energies of detached electron:
    (a) $E_{\rm cont}=50$~meV;
    (b) $E_{\rm cont}=40$~meV;
    (c) $E_{\rm cont}=30$~meV;
    (d) $E_{\rm cont}=24$~meV;
    (e) $E_{\rm cont}=20$~meV;
    (f) $E_{\rm cont}=10$~meV.
    }
\end{figure}

Matrix element of an electronic non-adiabatic coupling element calculated from Eq.~\ref{Eq:coupling} for six different energies of the ejected electron is shown in Fig.~\ref{fig:coupling_J0}. The kinetic energies of the ejected electron equal to the differences between the initial and final rovibronic states are in the range from 10 meV to 50 meV and are typical for the case under consideration. The mentioned electronic non-adiabatic couplings allow the calculation of the vibrational coupling element between selected states according to Eq.~\ref{Eq:vibronic coupling}. Assuming that the energy of the detached electron is in the range given above, and the maximal lifetime of the electron emission reaction is less than 10 s, which corresponds to the duration of the experiment reported by Anderson \emph{et al.}\cite{Anderson2020}, then the mean value of this coupling element between all possible rovibronic states of Ag$_2^-$ and Ag$_2$ for $J=0$ is 1.4$\cdot$10$^{-9}$ a.u. Averaging is performed by all possible pairs of vibrational levels, for which energy gaps are in the range related to the kinetic energy of the ejected electrons. As expected, the value of the vibrational coupling element is very small. In turn, the mean lifetime of the spontaneous electron emission (Ag$_2^-$ $\rightarrow$ Ag$_2$ + e$^-$) as estimated by Eq. (\ref{Eq:ad level width}), is around 3 seconds, and the shortest calculated lifetime is 262~ms.  

All this confirms the experimental finding that both processes (i.e. spontaneous electron emission and rotational predissociation) compete with each other in the same timescale and their duration can be quite long, in specific cases up to a few seconds.

\section{\label{sec:level4} Conclusions}

In reference to the recent experiment, we set ourselves the goal of creating a theoretical description of the two-channel decay of a diatomic silver anion molecule containing spontaneous emission of electrons and defragmentation. We calculated the electronic ground states of anion and neutral diatomic silver molecules. Systematic calculations revealed the Morse-like shape potentials in both cases, with a much more pronounced deviation for the attractive part of the anion potential. The map of the rovibrational levels of both systems shows the dependence of the decay channels on the rotational quantum number $J$ and the energy of the vibronic levels. This is broadly in line with the experimental conclusions. The region where both channels are open is relatively small. The highest value of $J$, for which both channels are still open called the critical $J_c$, is 253. In the case of $J$ higher than the critical $J_c$, only the defragmentation channel is open.

Electronic non-adiabatic coupling elements were calculated as a function of the kinetic energy of the ejected electron. Since then, there has been no crossing or avoided-crossing of the potential curves, its value is small, which makes the calculated mean value of the vibrational non-adiabatic coupling between all considered rovibrational levels very small, of the order of 10$^{-9}$ a.u. This means that the spontaneous electron emission process can be long and has a mean lifetime estimated by our approach of a few seconds, although the shortest lifetimes are on a millisecond scale (i.e. $\sim$250~ms).

In turn, the mean lifetime of the predissociation reaction calculated for the considered quasibond states is smaller compared to the electron autodetachment reaction and equals around 250~ms. It means that the fragmentation reaction is generally faster than the spontaneous electron emission reaction occurring even on pico- and nanosecond scale, but some overlap in time (starting with milliseconds) of both channels can be observed, leading to vie of these two decay pathways of the Ag$_2^-$ molecule. All the above confirms the experimental results on the timescale of reactions given by Anderson \emph{et al.}\cite{Anderson2020}.

\section*{Acknowledgements}

We acknowledge partial support from the COST Action “Attosecond Chemistry" (Grant No. CA18222).
This work was supported by computer grants
from the computer centres WCSS (Wroclawskie Centrum Sieciowo-Superkomputerowe, Politechnika Wroclawska) and CI TASK (Centrum Informatyczne Tr\'ojmiejskiej Akademickiej Sieci Komputerowej, Politechnika Gda\'nska).
J. F. thanks Nico Sanna for helpful discussions.

\section*{Data Availability}
The data that support the findings of this study are available
from the corresponding author upon reasonable request and are
openly available in MOST Wiedzy at https://mostwiedzy.pl/en/open-research-data-series/niema,202103311024423091181-0/catalog.\\

\bibliography{Ag2-.bib}

\end{document}